\documentclass{article}

\usepackage{arxiv}

\usepackage[utf8]{inputenc} 
\usepackage[T1]{fontenc}    
\usepackage{hyperref}       
\usepackage{url}            
\usepackage{booktabs}       
\usepackage{amsfonts}       
\usepackage{nicefrac}       
\usepackage{microtype}      
\usepackage{lipsum}		
\usepackage{graphicx}
\usepackage{natbib}
\usepackage{doi}
\usepackage{caption}
\usepackage{subcaption}
\usepackage{algorithm}
\usepackage{algorithmic}
\usepackage{xcolor}
\usepackage{varwidth}
\usepackage{cleveref}

\title{Comparing Human and AI Performance in Visual Storytelling through Creation of Comic Strips: A Case Study}

\author{
\href{https://orcid.org/0000-0001-8347-3038}{\includegraphics[scale=0.06]{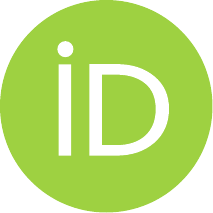}\hspace{1mm}Uğur Önal}\\
Artificial Intelligence and \\Data Engineering Department\\
Istanbul Technical University\\
Maslak, Istanbul, Turkey\\
	\texttt{onalug@itu.edu.tr} \\
    \And 
\href{https://orcid.org/0000-0003-2993-6681}{\includegraphics[scale=0.06]{orcid.pdf}\hspace{1mm}Sanem Sariel}\\
Artificial Intelligence and \\Data Engineering Department\\
Istanbul Technical University\\
Maslak, Istanbul, Turkey\\
	\texttt{sariel@itu.edu.tr} \\
 	\And 
 \href{https://orcid.org/0000-0002-1524-1646}{\includegraphics[scale=0.06]{orcid.pdf}\hspace{1mm}Metin Sezgin}\\
    Computer Science and \\Engineering Department\\
Koc University\\
Istanbul, Turkey\\	\texttt{mtsezgin@ku.edu.tr } \\
 	\And 
\href{https://orcid.org/0000-0003-3618-4166}{\includegraphics[scale=0.06]{orcid.pdf}\hspace{1mm}Ergun Akleman}\\
	Visual Computing \& Computational Media,\\ Joint with Computer Science and \\Engineering Department\\Texas A\&M University, \\College Station, TX, 77831\\
	\texttt{ergun@tamu.edu} 
 }

\renewcommand{\shorttitle}{Comparing Human and AI Performance in Visual Storytelling}

\hypersetup{
pdftitle={A template for the arxiv style},
pdfsubject={q-bio.NC, q-bio.QM},
pdfauthor={David S.~Hippocampus, Elias D.~Striatum},
pdfkeywords={First keyword, Second keyword, More},
}

\begin{document}

\maketitle

\begin{abstract}
This article presents a case study comparing the capabilities of humans and artificial intelligence (AI) for visual storytelling. We developed detailed instructions to recreate a three-panel Nancy cartoon strip by Ernie Bushmiller and provided them to both humans and AI systems. The human participants were 20-something students with basic artistic training but no experience or knowledge of this comic strip. The AI systems used were popular commercial models trained to draw and paint like artists, though their training sets may not necessarily include Bushmiller's work. Results showed that AI systems excel at mimicking professional art but struggle to create coherent visual stories. In contrast, humans proved highly adept at transforming instructions into meaningful visual narratives.
\end{abstract}

\section{Introduction and Motivation}
\label{Sec_Introduction}

Creating visually compelling images has traditionally required significant effort and expertise. Recently, AI-based drawing and painting systems have gained popularity, sparking controversies in the art community, such as a dispute in a cartoon competition \citep{cumhuriyet2024turhan,darroch2017netherland,kotbas2024yapay}.

While AI is argued to boost efficiency and create jobs requiring advanced skills \citep{trivedi2023should}, studies show people prefer human-made art for its emotional depth, narrative, and meaning \citep{bellaiche2023humans}. AI systems have notable limitations in graphic design \citep{sindhura2021virtues}. As a solution, hybrid intelligence has been proposed, combining human and AI capabilities to enhance performance, as seen in projects like celestial body classification via crowdsourcing \citep{kamar2012combining} and integrating expert knowledge with AI to improve results \citep{chang2017revolt,bansal2021does}.

This paper demonstrates that AI struggles to create nuanced visual images for effective storytelling, emphasizing the need for human expertise to craft meaningful visuals.

\subsection{Context and Motivation}
\label{Sec_ContextandMotivation}

\begin{figure}[hbtp]
\centering
\includegraphics[width=1.00\textwidth]{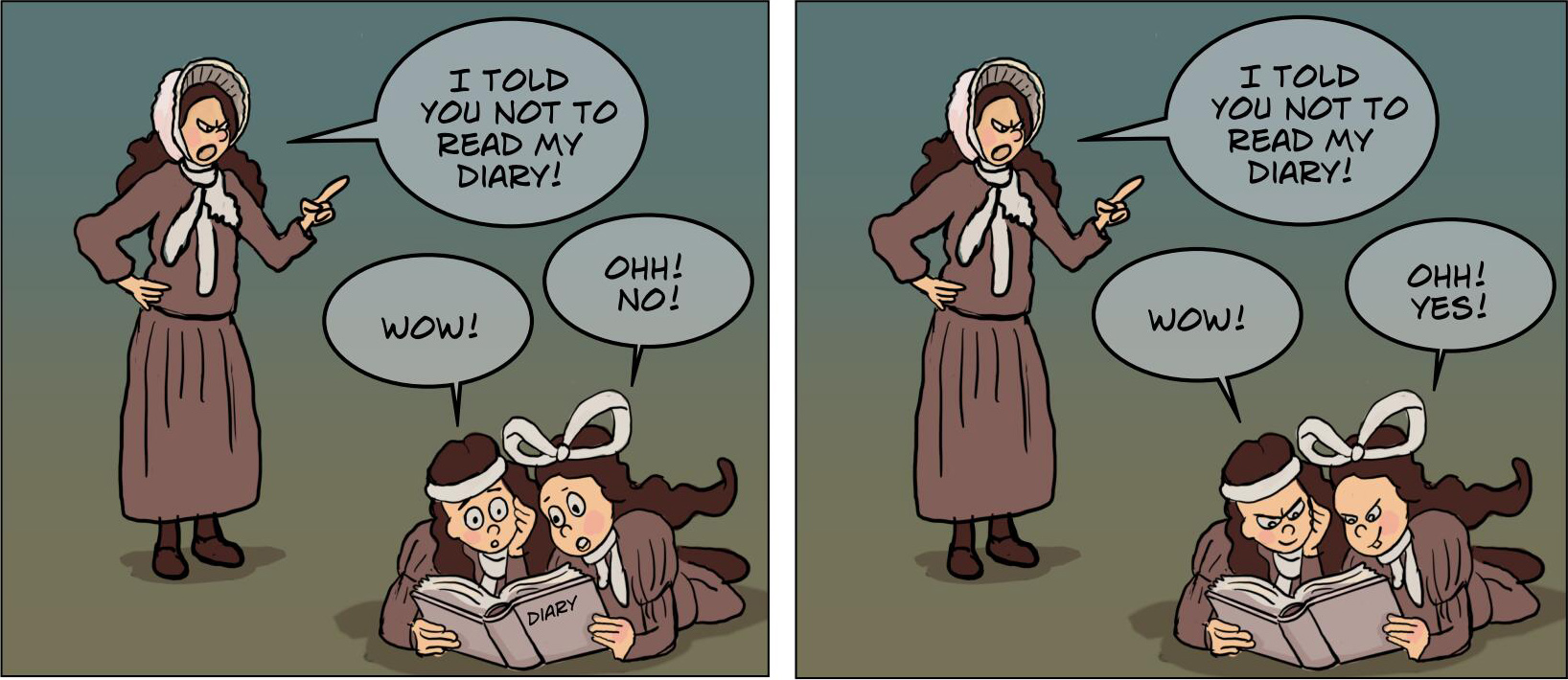}
    \caption{
An example highlights the role of subtle facial expressions in storytelling. In the left panel, the children appear surprised, likely finding the diary by accident, unaware of their sister's anger as they're engrossed in its contents. In contrast, the right panel shows mischievous children who deliberately searched for the diary, fully aware of their sister's angry gaze.  }
\label{fig_images_expressions}
\end{figure}

Effective visual storytelling is challenging, even for professionals. Subtle details play a crucial role, as illustrated in Figure~\ref{fig_images_expressions}, which compares two nearly identical cartoon panels on privacy published in IEEE Computer Magazine \citep{akleman2021privacy} \footnote{The left panel was used in the final cartoon.}. A minor change in the eyes transforms mischievous children into surprised ones, shifting the narrative from intentionally reading their sister's diary to an accidental discovery.

Effective storytelling relies on creating appropriate human affects, such as emotions, mood, or attachment. For AI, generating such affects is challenging as they involve complex expressive cues—facial, vocal, or gestural behaviors—learned by humans over decades of interactions \citep{vandenbos2007apa}. Subtle cues like body posture or gaze direction can significantly alter the perception of affects \citep{liu2012, akleman2015, dede2024power}. However, standard research on human affects focuses mainly on broad categories like facial, vocal, and gestural expressions \citep{ekman1979facial, ekman1997universal, ekman1999facial, russell2003facial}, struggling to classify subtle and uncommon expressions.

\subsection{Basis, Rationale and Contributions}
\label{Sec_BasisAndRationale}

\begin{figure}[hbtp]
\centering
\includegraphics[width=0.99\linewidth]{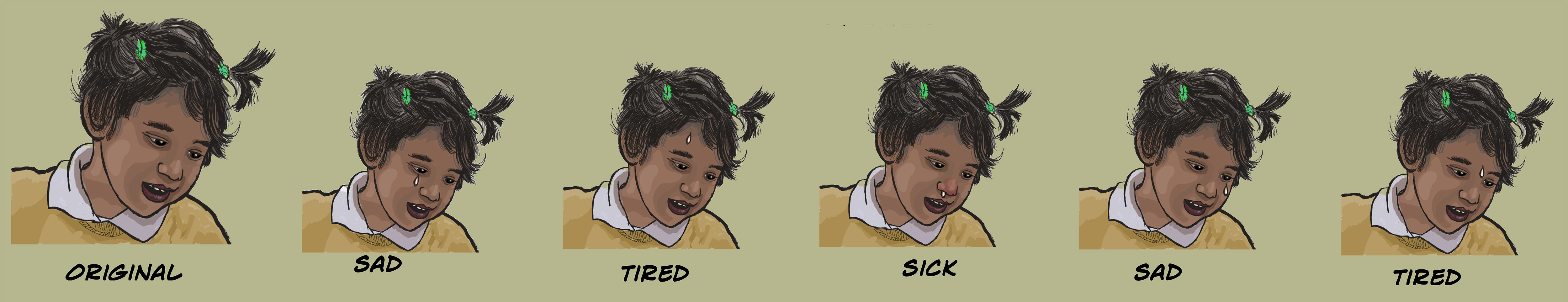}
    \caption{The position and orientation of small props like droplets can alter perceived expressions \citep{akleman2020}. Droplets near the eyes suggest sadness, on the forehead indicate fatigue, and under the nose imply sickness, even on a neutral face. More examples can be found in \citep{akleman2020}. }
\label{droplets/0}
\end{figure}

Creating human affects is challenging due to the subtle, context-dependent nature of expressive cues, which go beyond simple image and text inputs. Effective storytelling requires considering all sensory inputs within context, as small details like droplets or body posture significantly enhance expression \citep{akleman2020} (see Figure~\ref{droplets/0}). Subtle cues in storytelling, akin to micro-expressions in emotion recognition, are brief and difficult to synthesize or detect algorithmically \citep{merghani2018reviewfacialmicroexpressionsanalysis, 9915437}. While AI tools like facial image generators \citep{stylegan2} produce impressive visuals, minor inconsistencies can disrupt intended meanings \citep{FAN202297}.

Expert artists and animators excel at creating subtle, context-aware cues through practice, though few formally document their methods \citep{johnston1981illusion, mccloud1993understanding, blair1995cartoon, mccloud2006making, eisner2008comics, celik2011on}. Simplified user studies combined with expert knowledge could improve the understanding and application of these cues, aiding dataset creation \citep{dede2024power}.

Context is equally vital for understanding human interactions and storytelling. Defined in ubiquitous computing as "physical activity, location, and the psychophysiology and affective state of a person" \citep{bulling2011}, context shapes behavior and norms. Misunderstandings may arise from cultural differences, such as conflicting views on personal space. In human-computer interaction, context-aware systems enhance usability by adapting to specific situations, such as recognizing faster, louder speech in emergencies like house fires, where failure could have critical consequences.

In conclusion, understanding subtle expressions and their context is essential for compelling storytelling. Computational agents, such as AI systems, must analyze and adapt to their operational contexts. This study shows that AI systems still lag behind humans in generating nuanced expressions and contextual understanding necessary for effective storytelling.
\section{Process}
\label{Sec_Process}

Our process for comparing human and AI performance in visual storytelling through the creation of comic strips consists of four steps:
\begin{enumerate}
\item \textbf{Selection of a Comic Strip:} \\In this stage, the goal is to select well-known comic strips that are recognized and praised by visual storytelling experts for their effectiveness. 
\item \textbf{Creation of verbal instructions, i.e. prompts:}\\ This stage involves describing the chosen comic strip in as much detail as possible using only text. These descriptions, referred to as prompts in the context of artificial intelligence, are essentially verbal instructions for completing a task. Such verbal instructions have long been used by humans to describe tasks.
\item \textbf{Creation of Comic Strips by using verbal instructions:} \\In this stage, the same verbal instructions are sent to both humans and various AI systems to produce corresponding comic strips.
\item \textbf{Analyzing comic strips in terms of their quality:} \\In this final stage, the quality of the comic strips is analyzed and compared.
\end{enumerate}
This process is general and can be repeated with many different comic strips. However, in this paper, we focus on a single strip, as it is well-known for its quality.

\subsection{Selection of Comic Strip}

\begin{figure}[hbtp]
\centering
\includegraphics[width=0.99\textwidth]{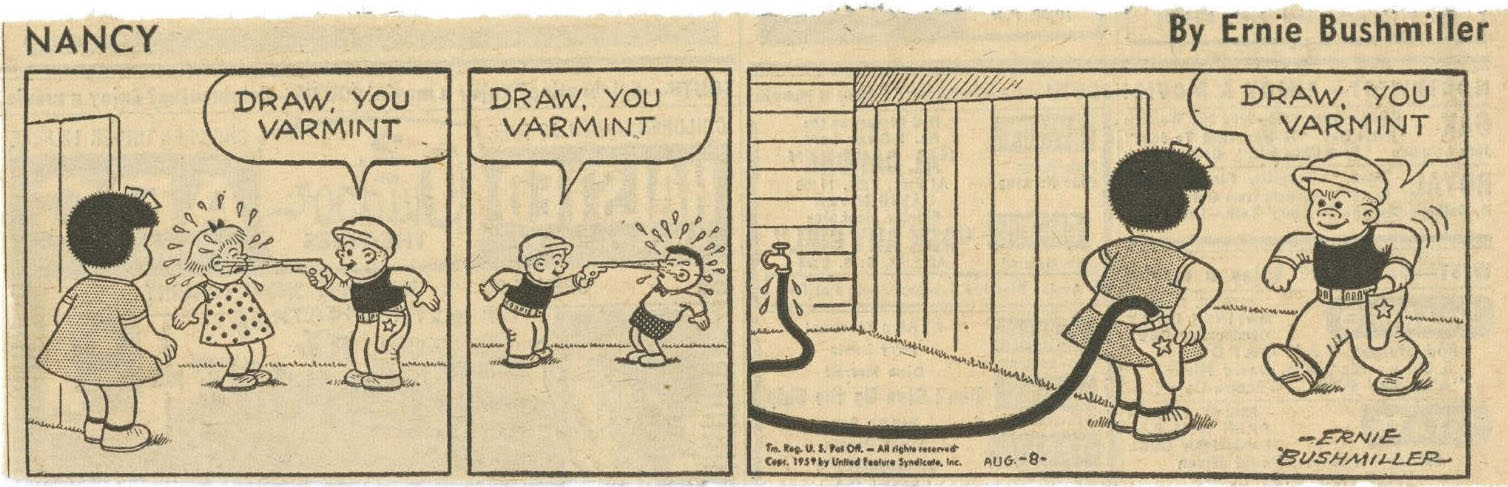}
    \caption{The original "How to Read Nancy" comic strip drawn by Ernie Bushmiller that the prompt is derived from. To the best of our knowledge, this particular strip is in the public domain. While the copyrights for many Nancy comic strips are held by Andrews McMeel Syndication, this specific strip does not appear in their published list of copyrighted works. We contacted the syndicate on two separate occasions, and they did not assert copyright ownership over this strip.}
\label{fig_images_Nancy_Original}
\end{figure}

To compare the creative capabilities of human and artificial intelligence, we used a prompt based on a famous Ernie Bushmiller cartoon strip (see Figure \ref{fig_images_Nancy_Original}). This strip was chosen because it is extensively analyzed in Paul Karasik and Mark Newgarden’s Eisner Award-winning book, How to Read Nancy: The Elements of Comics in Three Easy Panels \citep{newgarden1988how,newgarden2017how}. The book, recognized as Best Comics-Related Book in 2018, highlights the strip's artistic significance, arguing that Nancy, often dismissed as simple, is a cornerstone of comic art.

Karasik and Newgarden dissect this Nancy strip panel by panel, revealing how lines, shapes, timing, and space contribute to its storytelling and humor. They argue that Bushmiller's simplicity masks his precision and mastery. The authors highlight universal comic principles like pacing, visual economy, and text-image integration, showcasing this strip as a model of effective comic art.

They argue that the strip's humor is crafted with mathematical precision, relying on perfect element placement and timing. Bushmiller's methods offer valuable lessons for aspiring cartoonists, demonstrating the storytelling power of visual art and highlighting often-overlooked nuances in comic storytelling.

This strip, celebrated for its wit and depth, is ideal for testing how human or artificial intelligence replicates the mechanics of comics. Its visual economy suggests that transforming illustrations into a cohesive strip reveals the complexity of a specific type of intelligence.

\begin{table}[hbtp]
  \framebox{%
   \begin{varwidth}{\textwidth}
    \begin{quote}
    We want you to create a three-panel comic strip. Each panel will be positioned side by side. There will be two main characters in the comic strip: "X" and "Y." Instructions for drawing the characters and panels are provided below:
    \end{quote}
    \begin{quote}
    \textbf{Character Instructions:}
    \begin{itemize}
    \item \textbf{Character X:} This character is a 10-year-old girl wearing a skirt and a simple T-shirt, with the T-shirt tucked in. She has short, black, curly hair and is slightly overweight.
    \item \textbf{Character Y:} This character is a 10-year-old boy wearing a baseball hat. He wears shorts and a simple T-shirt and carries a water gun, which he uses to wet everyone around him.
    \end{itemize}
    \end{quote}
    \begin{quote}
    \textbf{Instructions for Panels:}
    \begin{itemize}
    \item \textbf{Panel 1:} The character X is positioned on the left side of the panel, seen from behind, observing Y as he sprays a girl with his water gun. The character X looks concerned about the situation.
    \item \textbf{Panel 2:} The character Y is in the center of the action, spraying another boy with his water gun. The character X is not visible in this panel.
    \item \textbf{Panel 3:}  The character X reappears on the left side, again seen from behind. This time, she has a water hose hidden behind her, ready to retaliate against the character Y. On the right side, The character Y is walking toward the character X with a smirk on his face, oblivious to what’s coming.
    \end{itemize}
    \end{quote}
   \end{varwidth}
}
\caption{Instructions that are used to create the comic strips.}
\label{Prompt}
\end{table}

\subsection{Prompt Creation}

Reproducing this particular strip can offer fresh insights into the intelligence required to create the simplest-seeming works. We also expected that it would be possible to evaluate the importance of meticulous craftsmanship provided by Artificial Intelligence in terms of the creation of visual stories.

In the prompt, we specifically converted the names of the characters to X and Y to prevent AI tools from being influenced by character names if they exist in their database and to ensure human artists are not biased by prior knowledge of the ``How to Read Nancy" comics. Additionally, some AI tools impose a character limit on the prompts they process. You can see the instructions that are used to create the comic strips in Table \ref{Prompt}.

\begin{figure}[hbtp]
    \centering
        \begin{subfigure}[t]{0.99\textwidth}
    \includegraphics[width=0.99\textwidth]{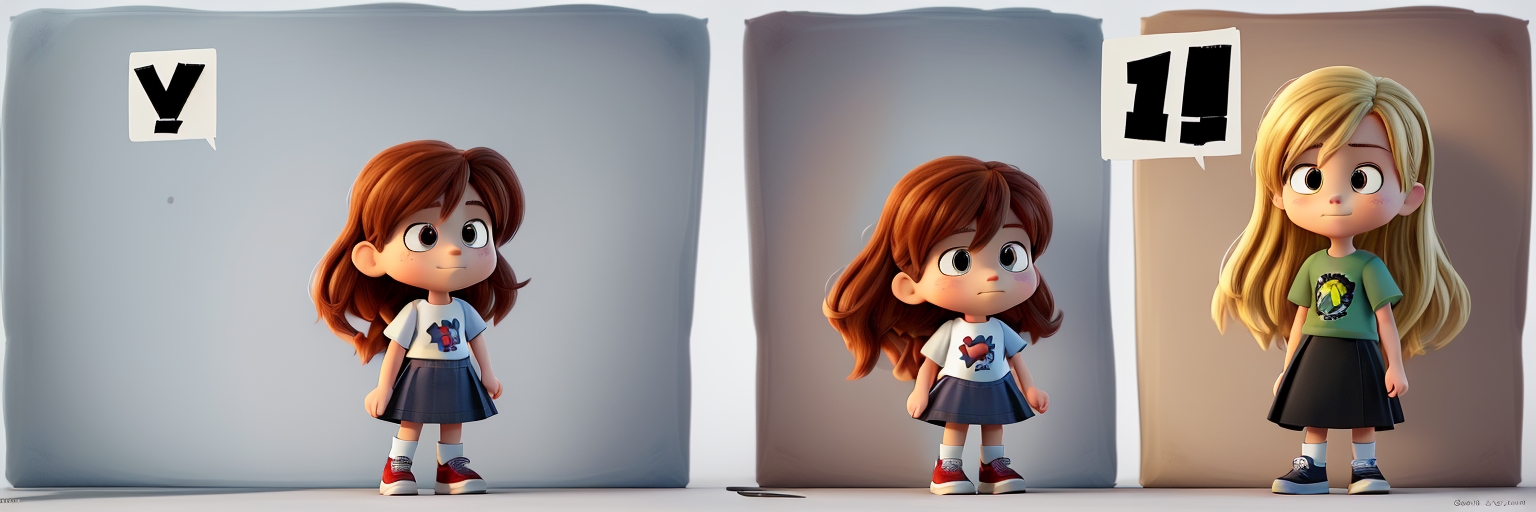}
    \caption{An example "comic" created by free model in ``3D Animation Style created by Leonardo.Ai\citep{leonardoai}.}
    \label{fig_images_3DAnimationStyle_None_0}
    \end{subfigure}
    \hfill
    \begin{subfigure}[t]{0.99\textwidth}
    \centering
    \includegraphics[width=0.99\textwidth]{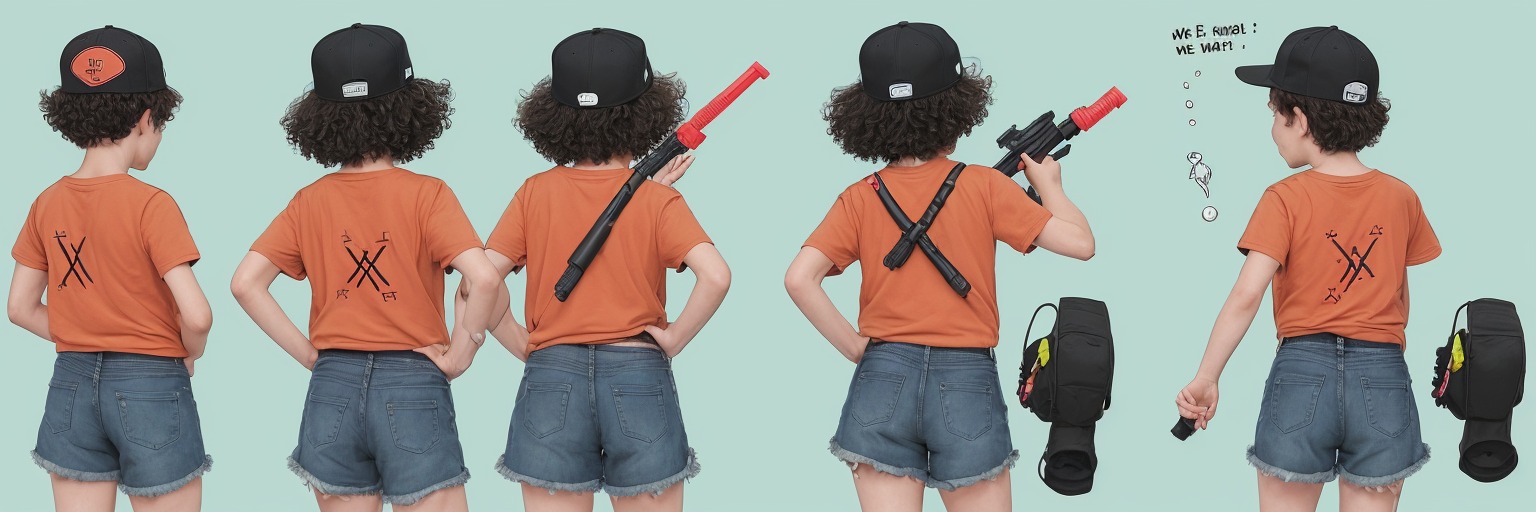}
    \caption{An example "cartoon" created by custom-trained model ``realisticVision'' of Stable Baselines.}
    \label{fig_images_StableBaselines_realisticVision_1}
    \end{subfigure}
    \hfill
    \begin{subfigure}[t]{0.99\textwidth}
    \centering
        \fbox{\includegraphics[height=0.2\textheight]{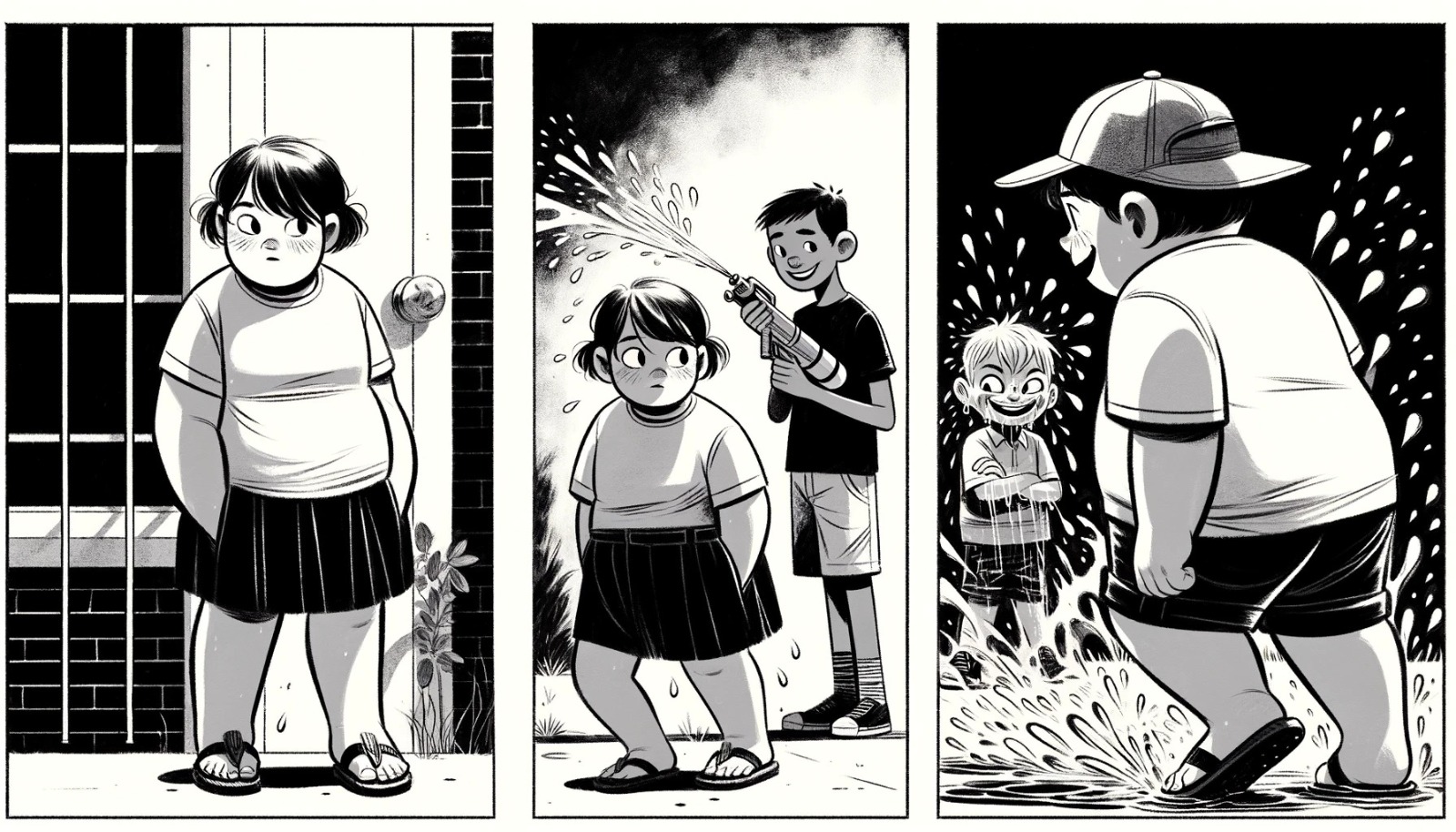}}
        \fbox{\includegraphics[height=0.2\textheight]{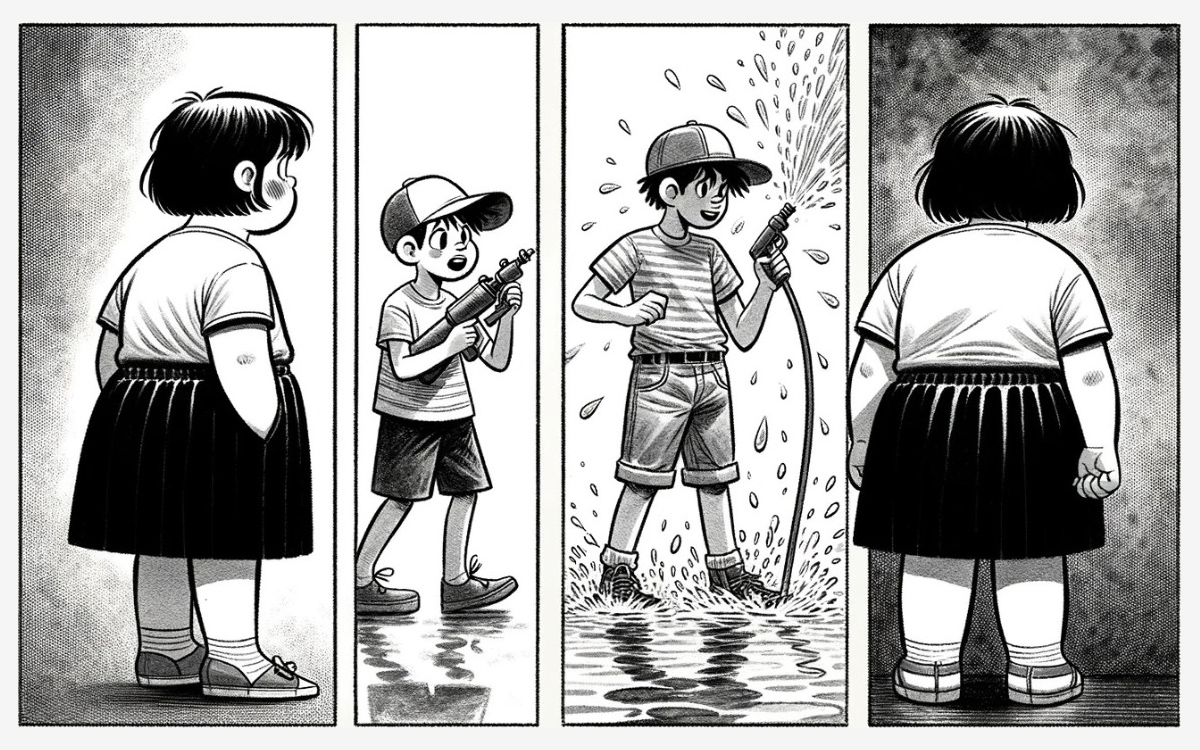}}
    \caption{Two examples created by OpenAI's Dall-E using ChatGPT.}
    \label{fig_images_Dall-E}
        \end{subfigure}
    \caption{The four "comic strips" created by AI using the prompt.}
    \label{fig_images_AI}
\end{figure}

\subsection{Creation of Comic Strips: AI vs. Humans}

After creating the prompt, it was given to three 20-something
students with basic artistic training but no experience or knowledge of this comic strip and various AI tools to compare their ability to visualize the instructions. A large set of outputs was generated using AI, but we present only the four best examples here, as shown in Figure~\ref{fig_images_AI}. The humans, who were unfamiliar with the comic strip, produced three strips, as shown in Figure \ref{fig_images_Human}.

\subsection{Analysis of Comic Strips: AI vs. Humans}

AI systems demonstrated strengths in generative image creation, especially in character visualization, producing visually accurate images based on prompts. Custom-trained AI models and tools yielded superior visual quality as shown in Figure~\ref{fig_images_AI}, showing potential when high-end resources are available. However, AI struggled with sequential contexts, such as comic strips, where its limitations overshadowed its strengths.

\begin{figure}[hbtp]
    \centering
        \begin{subfigure}[t]{0.99\textwidth}
        \includegraphics[width=0.99\textwidth]{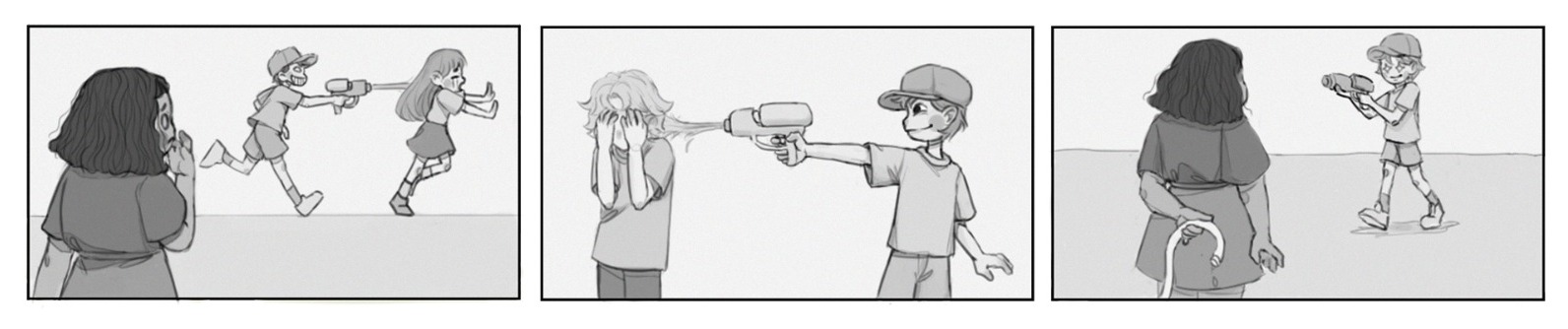}
        \caption{The comic strip drawn by Feyza Güreli.}
        \label{fig_images_Human_1}
    \end{subfigure}
    \hfill
    \begin{subfigure}[t]{0.99\textwidth}
    \includegraphics[width=0.99\textwidth]{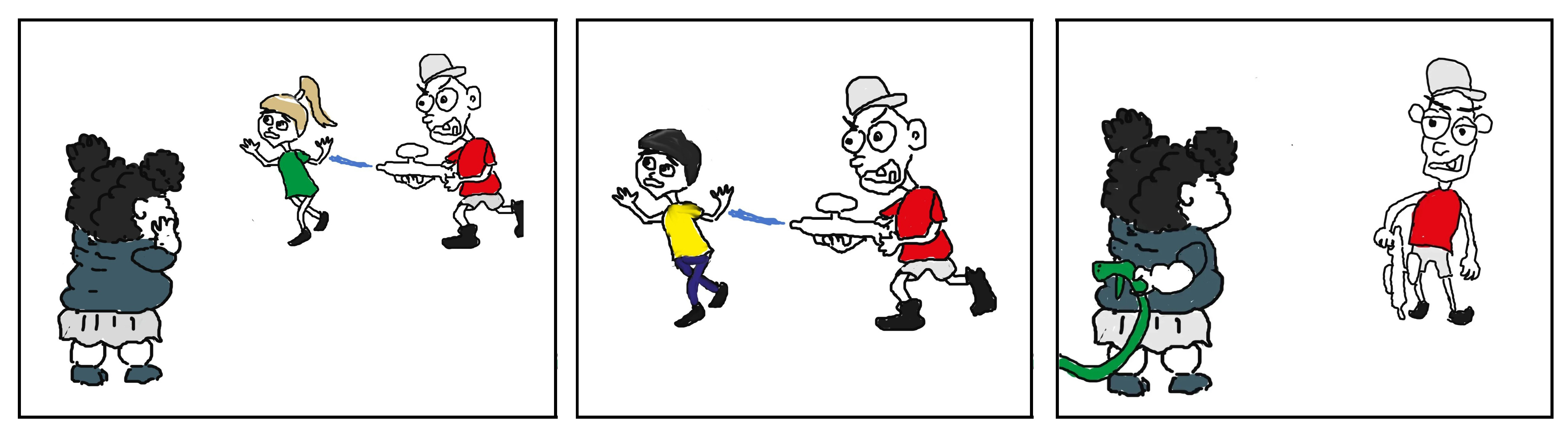}
        \caption{The comic strip drawn by Mücahit Morcol.}
        \label{fig_images_Human_2}
    \end{subfigure}
    \hfill
    \begin{subfigure}[t]{0.99\textwidth}
        \includegraphics[width=0.99\textwidth]{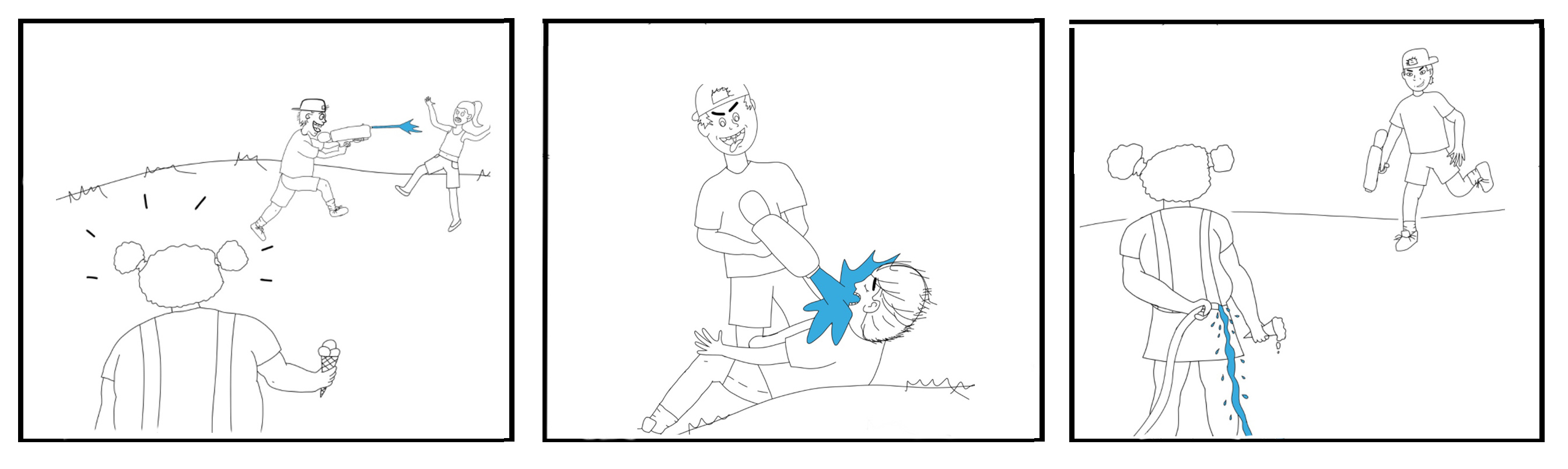}
        \caption{The comic strip drawn by Abdullah Raşid Gün.}
        \label{fig_images_Human_0}
    \end{subfigure}
    \caption{The three comic strips created by 20-something
students with basic artistic training but no experience or knowledge of this comic strip only based on the prompt provided to them.}
    \label{fig_images_Human}
\end{figure}

In contrast, human-generated comics consistently captured the prompt's narrative, showing a bully with a water gun disturbing others and a girl secretly preparing a stronger weapon \ref{fig_images_Human}. These works incorporated subtle nuances that enriched the story, highlighting a key advantage of human intelligence over AI. AI often overlooked these minor yet impactful details essential for enhancing storytelling depth and quality.

\section{Conclusion and Future Work}
\label{Sec_Discussion}

This work highlights a significant distinction between the intelligence demonstrated by AI and humans in creative tasks. AI-generated artworks often exhibit impressive detail and technical precision, yet they lack the fundamental elements required for effective storytelling. In contrast, humans not only understand instructions well but also excel at reproducing the subtle details necessary for visual narratives.

In summary, our findings show that while AI systems are proficient at mimicking professional drawing and painting styles, they fall short in crafting coherent visual stories. Humans, on the other hand, demonstrate exceptional capability in transforming well-defined instructions into compelling and meaningful visual narratives.

There is a need for similar studies, using the same or different comic strips, to track the progress of AI systems in visual storytelling.

\subsection{Disclaimer}

This study was conducted in 2024. A new version of ChatGPT, released in March 2025, began generating significantly improved cartoon strips from the same prompts, although some minor issues with facial expressions remain as shown in Figure 6. This development suggests that much stronger support for visual storytelling can be expected in the near future.

\begin{figure}[hbtp]
\centering
\includegraphics[width=0.99\textwidth]{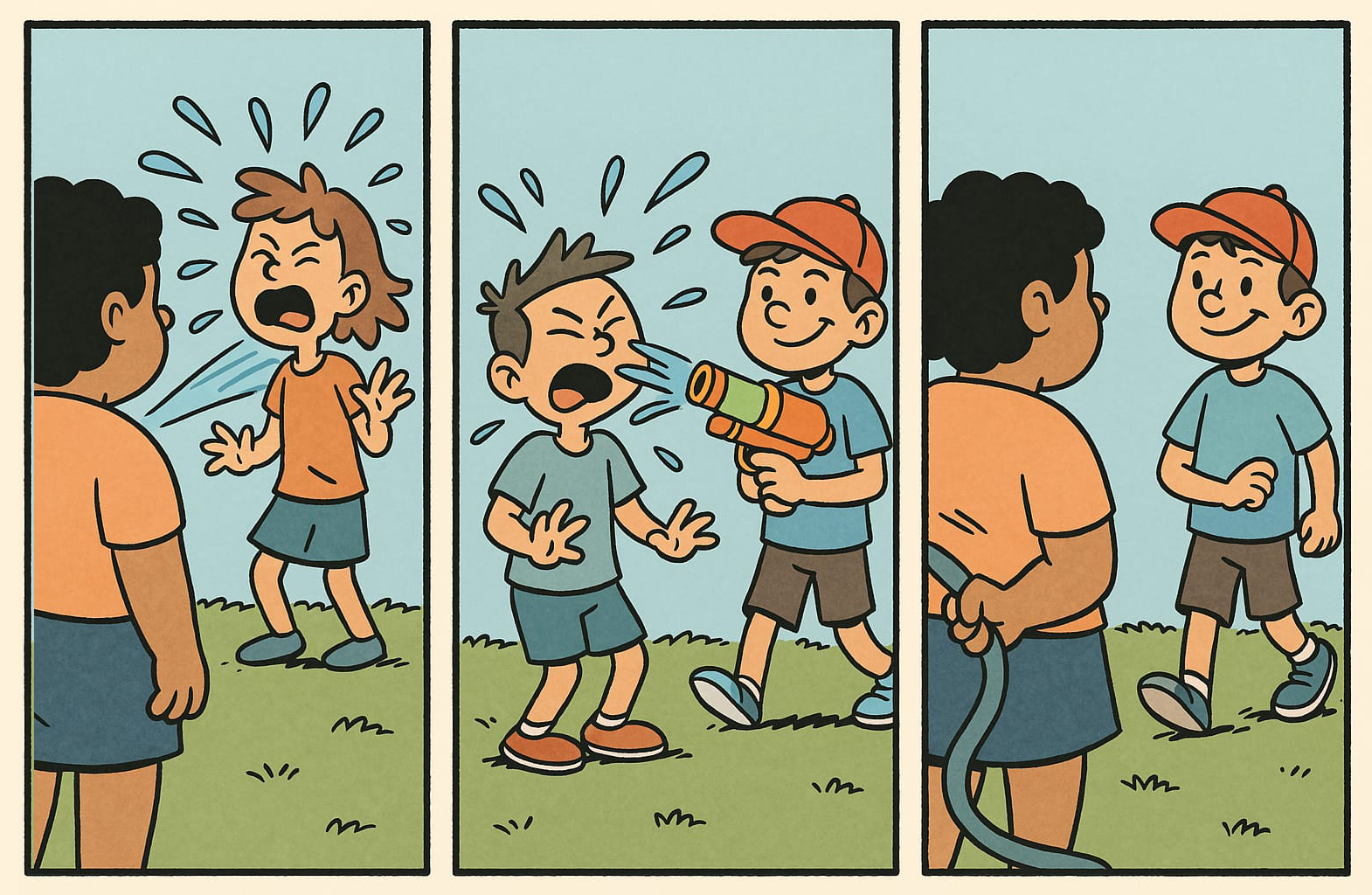}
    \caption{This is an example of the output generated using the prompt in the table with the new version of ChatGPT released in March 2025.}
\label{fig_chatGPT2025}
\end{figure}

\subsection{Acknowledgment}
\label{Sec_Acknowledgment}

We extend our heartfelt gratitude to Abdullah Raşid Gün, Feyza Güreli, and Mücahit Morcol for their invaluable contributions in creating the human sketches based on the given prompts. We are also deeply thankful to Doğukan Kıvrak and Sena Erokyar for their support with paid and custom-trained AI models and for sharing their final outputs.

\bibliographystyle{apalike}
\bibliography{references}

\end{document}